\documentclass[final]{elsart5p}

\usepackage{graphicx}

\usepackage{amssymb}
\usepackage[dvips]{color}

\begin{document}

\begin{frontmatter}


\title{Defects-driven appearance of half-metallic ferrimagnetism in Co-Mn--based Heusler alloys}

\author[Gebze]{K. \"Ozdo\~gan}\ead{kozdogan@gyte.edu.tr}
\author[Patras]{I. Galanakis\corauthref{cor}}\ead{galanakis@upatras.gr}
\author[Julich,Fatih]{E. \c Sa\c s\i o\~glu}\ead{e.sasioglu@fz-juelich.de}
\author[Gebze]{B. Akta\c s}
\address[Gebze]{Department of Physics, Gebze Institute of Technology,
Gebze, 41400, Kocaeli, Turkey}
\address[Patras]{Department of Materials Science, School of Natural
  Sciences, University of Patras,  GR-26504 Patra, Greece}
\address[Julich]{Institut f\"ur Festk\"orperforschung, Forschungszentrum
J\"ulich, D-52425 J\"ulich, Germany}
\address[Fatih]{Fatih University,
Physics Department, 34500, B\" uy\" uk\c cekmece,  \.{I}stanbul,
Turkey} \corauth[cor]{Corresponding author. Phone +30-2610-969925,
Fax +30-2610-969368}

\begin{abstract}

 Half-metallic ferromagnetic full-Heusler alloys containing
Co and Mn, having the formula Co$_2$MnZ where Z a sp element, are
among the most studied Heusler alloys due to their stable
ferromagnetism and the high Curie temperatures which they present.
Using state-of-the-art electronic structure calculations we show
that when Mn atoms migrate to sites occupied in the perfect alloys
by Co, these Mn atoms have spin moments antiparallel to the other
transition metal atoms. The ferrimagnetic compounds, which result
from this procedure, keep the half-metallic character of the
parent compounds and the large exchange-splitting of the Mn
impurities atoms only marginally affects the width of the gap in
the minority-spin band. The case of [Co$_{1-x}$Mn$_x$]$_2$MnSi is
of particular interest since Mn$_3$Si is known to crystallize in
the Heusler $L2_1$ lattice structure of Co$_2$MnZ compounds.
Robust half-metallic ferrimagnets are highly desirable for
realistic applications since they lead to smaller energy losses
due to the lower external magnetic fields created with respect to
their ferromagnetic counterparts.
\end{abstract}

\begin{keyword}
A.magnetically ordered materials

\PACS 75.47.Np \sep 75.50.Cc \sep 75.30.Et
\end{keyword}
\end{frontmatter}

\section{Introduction}

The emergence of spintronics, also known as magnetoelectronics,
the last decade brought in the center of scientific research new
properties and materials which had received little attention in
the past \cite{Zutic}. An important class of materials which are
at the present under intense study are the so-called half-metals
\cite{book}. These materials are hybrids between metals and
semiconductors or insulators, presenting metallic behavior for one
spin-band and semiconducting for the other, and thus overall they
are either ferro- or ferrimagnets with perfect spin-polarization
at the Fermi level \cite{Reviews}. de Groot and his collaborators
in a pioneering paper published in 1983 predicted the existence of
half-metallicity in the case of the intermetallic Heusler alloy
NiMnSb \cite{deGroot}. Since then several half-metallic
ferromagnetic materials have been initially predicted by
theoretical ab-initio calculations and after synthesized
experimentally.

Half-metallic Heusler alloys are expected to play a key role in
realistic applications due to their very high Curie temperatures
and their structural similarity to the widely used binary
semiconductors crystallizing in the zinc-blende structure
\cite{Landolt}. Initially the research was focused on the
so-called half- or semi-Heusler compounds like NiMnSb
\cite{Reviews,Gala-Half} but lately the interest has been shifted
to the so called full-Heusler compounds and mainly to the ones
containing Co, like Co$_2$MnAl or Co$_2$MnGe, which were already
known since early 70's \cite{Webster}. In early 90's it was argued
in two papers by a japanese group that they should be
half-metallic ferromagnets \cite{Ishida-Fujii} and
first-principles calculations by Picozzi et al \cite{Picozzi} and
Galanakis et al \cite{Gala-Full} in 2002 confirmed their
predictions. These papers acted as an inspiration to
experimentalists who devote a constantly increasing number of
publications to the study of their properties (see Refs.
\cite{book,Reviews,Westerholt,Elmers,Marukame,Kelekar,Reiss,Sakuraba,magnetism,superlattices,transport,devices}
for references to some of these studies). Both the origin of
ferromagnetism \cite{Sasioglu} and half-metallicity
\cite{Gala-Full} in these compounds are well understood.

Although the research on half-metallic ferromagnets is intense,
the ideal case would be a half-metallic antiferromagnet, also
known as fully-compensated ferrimagnet  \cite{Leuken}, since such
a compound would not give rise to stray flux and thus would lead
to smaller energy consumption in devices.  In the absence of such
an ideal compound, half-metallic ferrimagnetism is highly
desirable  since such compounds would yield lower total spin
moments than their ferromagnetic counterparts. Some perfect
Heusler compounds like FeMnSb \cite{Groot2} and Mn$_2$VAl
 \cite{Kemal}  are predicted to present half-metallicity in conjunction with ferrimagnetism.
 Recently other routes  to half-metallic ferrimagnetism have been
studied like the doping of diluted magnetic semiconductors
\cite{Akai}, and the inclusion of defects in Cr pnictides
\cite{Galanakis-RC} and Co$_2$CrAl(or Si) full-Heusler alloys
\cite{PSS-RRL}.

In this communication we will study the appearance of
defects-driven ferrimagnetism in the popular Co$_2$MnZ compounds
where Z stands for Al, Ga, Si, Ge or Sn. Most of the theoretical
studies on these compounds either are devoted to the perfect
compounds \cite{Gala-Full,Podlucky,Richter} or concern disorder
between the Mn and the sp atoms or doping of the Mn sites (see
Refs. \cite{JAP,APL,different} and references therein). When a
surplus of Mn atoms is created occupying the sites corresponding
to Co atoms at the perfect compounds,  [Co$_{1-x}$Mn$_x$]$_2$MnZ
alloys, these Mn impurity atoms are shown to be
antiferromagnetically coupled to the other Co and Mn atoms at the
perfect lattice sites. The resulting ferrimagnetic compounds keep
the half-metallic character of the perfect parent alloys.
Interestingly due to the very high exchange splitting presented by
the Mn impurity atoms, the width of the gap is only marginally
affected contrary to the Co$_2$CrAl and Co$_2$CrSi where the
creation of Cr antisites almost vanished the gap \cite{PSS-RRL}.
Thus the defects-driven half-metallic ferrimagnetism presented in
this communication is of particular interest for realistic
applications. Special attention should be given to
[Co$_{1-x}$Mn$_x$]$_2$MnSi alloys since Mn$_3$Si is known
experimentally to crystallize in the Heusler $L2_1$ lattice
structure of Co$_2$MnZ compounds with two equivalent type of Mn
atoms in the unit cell \cite{Mn3Si}. On the contrary Mn$_3$Ge and
Mn$_3$Sn compounds crystallize in an hexagonal structure
\cite{Mn3Ge} while no information is available on the Mn$_3$Al or
Mn$_3$Ga compounds. We should also note here that Picozzi and
collaborators studied in Ref. \cite{Picozzi04} the case of a
single Mn antisite using a supercell structure in the case of
Co$_2$MnSi and Co$_2$MnGe compounds. They had actually found that
the Mn impurity atoms has a spin moment antiparallel to the other
transition metal atoms but they had not considered the case of
extensive defects or studied in detail the observed behavior in
their calculations.

We employed the full--potential nonorthogonal local--orbital
minimum--basis band structure scheme (FPLO) to perform the
electronic structure calculations and the coherent potential
approximation (CPA) to simulate the creation of the Mn antisites
\cite{koepernik}. We used the scalar relativistic formulation and
thus the spin-orbit coupling was not taken into account. The
exchange--correlation potential was chosen in the local spin
density approximation (LSDA). The self-consistent potentials were
calculated on a $20\times20\times20$ \textbf{k}-mesh in the
Brillouin zone, which corresponds to 256 k points in the
irreducible Brillouin zone. The lattice constants were the
experimental ones, 0.5756 nm for Co$_2$MnAl, 0.577 nm for
Co$_2$MnGa,  0.565 nm for Co$_2$MnSi, 0.574 for Co$_2$MnGe and
0.598 nm for Co$_2$MnSn \cite{Landolt}, and were kept constant
upon the creation of defects.

\begin{figure}
\begin{center}
\includegraphics[scale=0.6]{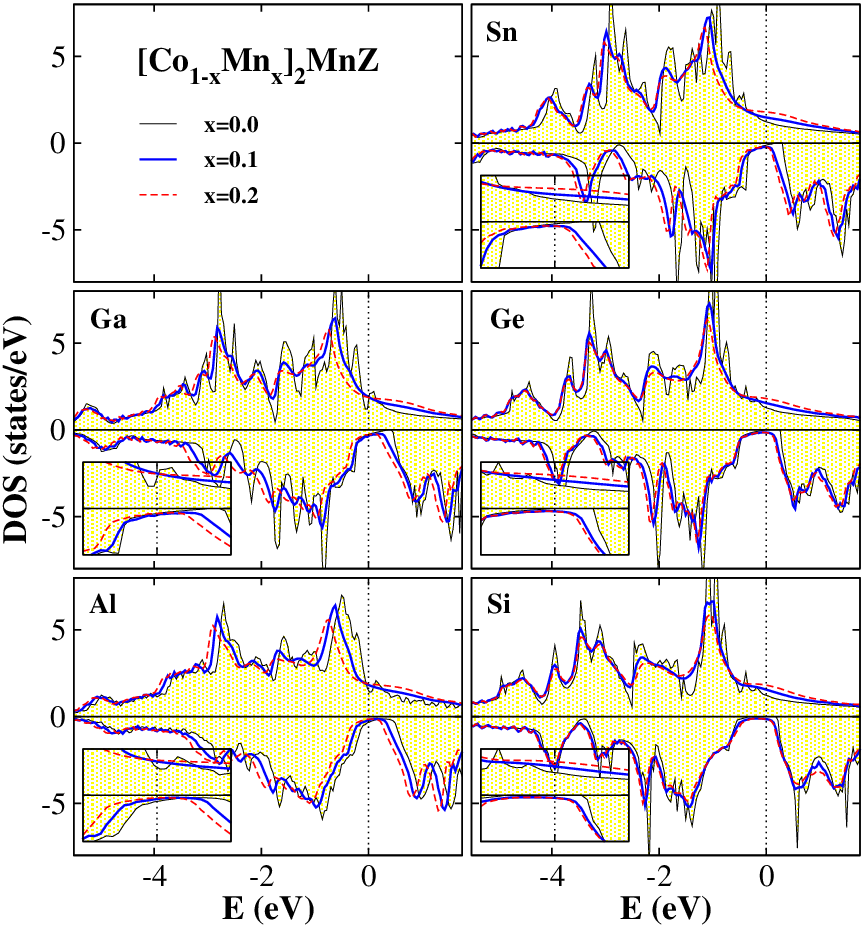}
\end{center}
\caption{
 Total density of states (DOS) for the [Co$_{1-x}$Mn$_{x}$]$_2$MnZ alloys as a function of the
concentration $x$ : we denote $x=0$ with the solid grey line with
shaded region, $x=0.1$ with a solid thick blue line and $x=0.2$
with a dashed red line. The Fermi level has been chosen as the
zero of the energy axis, and positive values of DOS correspond to
the spin-up (majority) electrons while negative values correspond
to the spin-down (minority) electrons. In the insets we have blown
up the region around the Fermi level.} \label{fig1}
\end{figure}

\section{Results and discussion}

We have performed calculations for the [Co$_{1-x}$Mn$_x$]$_2$MnZ
compounds varying the sp atom, Z, which is one of Al, Ga, Si, Ge
or Sn. We have taken into account five different values for the
concentration $x$; $x$= 0, 0.025, 0.05, 0.1, 0.2. In Fig.
\ref{fig1} we have drawn the total density of states (DOS) for all
five families of compounds under study and for three different
values of the concentration $x$ : the perfect compounds ($x$=0)
and for two cases with defects, $x$= 0.1 and 0.2. In the case of
the perfect Co$_2$MnSi and Co$_2$MnGe compounds there is a real
gap in the minority spin band and the Fermi level falls within
this gap and these compounds are perfect half-metals. The other
three perfect compounds -Co$_2$MnAl, Co$_2$MnGa and Co$_2$MnSn-
present in reality a region of tiny minority-spin DOS instead of a
real gap, but the spin-polarization at the Fermi level only
marginally deviates from the ideal 100\% and these compounds can
be also considered as half-metals. These results agree with
previous electronic structure calculations on these compounds
\cite{Picozzi,Gala-Full,Richter,JAP,APL} and are confirmed by the
calculated total spin moments as discussed latter in the text.
When we create a surplus of Mn atoms which migrate at sites
occupied by Co atoms in the perfect alloys, the gap persists and
all compounds retain their half-metallic character. This is
clearly seen in the insets where we have blown the region around
the Fermi level. The most interesting case is the two compounds
which presented a real gap, Co$_2$MnGe and mainly Co$_2$MnSi. The
creation of Mn antisites, especially in the Si-case, does not
alter the width of the gap and half-metallicity is extremely
robust in these alloys with respect to the creation of Mn
antisites. We will explain this behavior latter in the text, when
we will discuss the atom-resolved DOS.

\begin{figure}
\begin{center}
\includegraphics[scale=0.6]{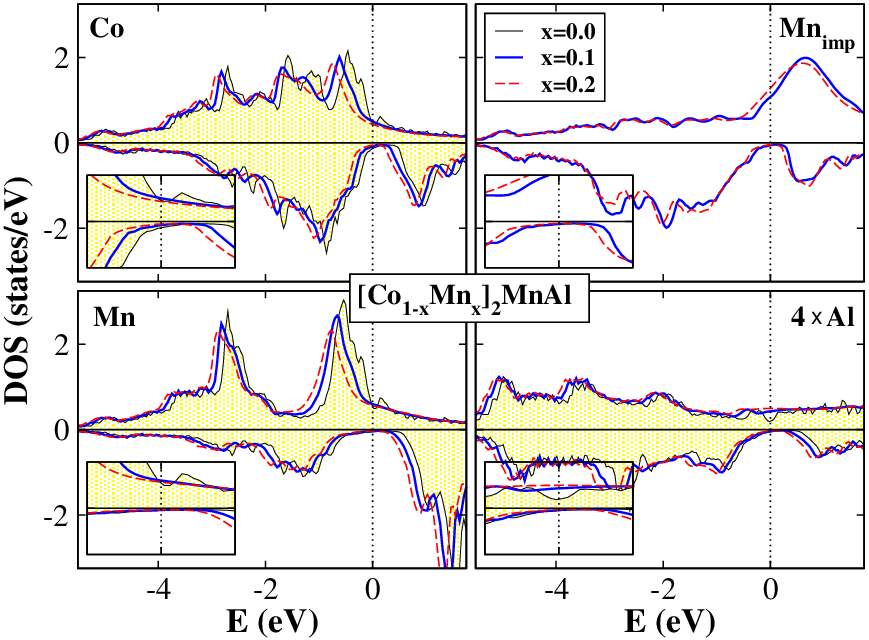}
\includegraphics[scale=0.6]{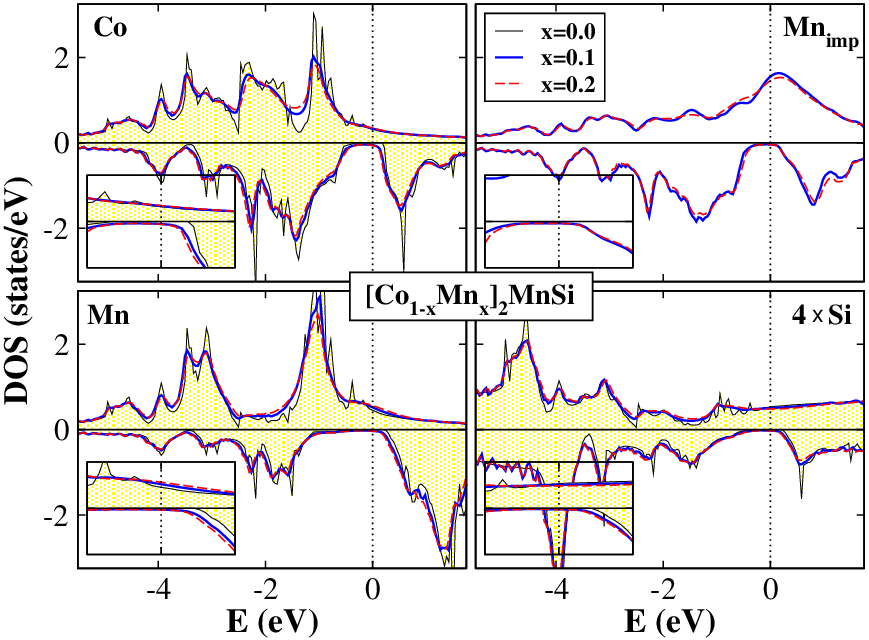}
\end{center}
\caption{
 Atom-resolved DOS as a function of the
concentration $x$ for the [Co$_{1-x}$Mn$_{x}$]$_2$MnAl (upper
panel) and [Co$_{1-x}$Mn$_{x}$]$_2$MnSi (lower panel) compounds as
a function of the concentration $x$. We have used the same
notation as in Fig. \ref{fig1}. Note that the atomic DOS's have
been scaled to one atom. The DOS for Al and Si has been multiplied
by a factor of four. Details are similar to Fig. \ref{fig1}. }
\label{fig2}
\end{figure}

In Fig. \ref{fig2} we have drawn the atom resolved DOS for the Al-
and Si-based compounds presented in Fig. \ref{fig1}. In the
perfect compounds, $x$=0, the Co and Mn atoms form a common
majority-spin band while the minority occupied states are mainly
of Co character. Around the minority-spin band-gap the states are
mainly of Co character confirming the results in Ref.
\cite{Gala-Full}. The large exchange splitting between the
occupied majority-spin states of Mn atoms and its unoccupied
minority states favors the creation of the gap and the appearance
of half-metallicity being also responsible for the large spin
moments at the Mn sites (see next paragraph). The DOS of the sp
atoms is very small with respect to the transition metal atoms and
thus we have multiplied it by four to make it visible. Al or Si
states around the Fermi level are of $p$ character and they play a
central role in the exact position of the Fermi level within the
gap (see Ref. \cite{Reviews} for an extended discussion of the
role of the sp atoms). When we substitute Si for Al, we increase
the total number of valence electrons in the unit cell by one and
this extra electron does not occupy states of the sp atom, which
are deep in energy, but rather states of the transition metal
atoms \cite{Gala-Full} provoking small changes in the DOS of the
Co and Mn atoms. When we create Mn antisites, the DOS of both Co
and Mn atoms at the perfect sites does not significantly change as
can be easily observed in Fig. \ref{fig2} and thus they only
marginally affect the half-metallicity. The interesting phenomenon
is the behavior of the Mn impurities atoms sitting at the
antisites. Now the major weight of the occupied states corresponds
to spin-down states and we expect their spin moments to be
antiparallel to the ones of the Co and Mn atoms at the perfect
sites. Moreover the large exchange splitting of the Mn atoms
ensures a large gap in the spin-down band and the robust character
of the half-metallicity. The shape of the DOS of the Mn impurity
atoms at the antisites is similar to the one of the Cr impurity
atoms in Co$_2$CrAl and Co$_2$CrSi \cite{PSS-RRL}. The main
advantage of the Mn-based compounds is that when Cr substitutes
Mn, the smaller exchange splitting of the Cr atoms leads to huge
narrowing of the band-gap which shrinks to an almost zero value
\cite{PSS-RRL}. Thus the Mn compounds have a clear advantage for
realistic applications.

\begin{table}
\caption{ Atom-resolved spin magnetic moments (in $\mu_B$) for the
[Co$_{1-x}$Mn$_x$]$_2$MnZ compounds (moments have been scaled to
one atom), where Z is Al or Ga. The last column is the total spin
moment (Total) in the unit cell calculated as $2\times
[(1-x)*m^{Co}+x*m^{Mn(imp)}]+m^{Mn}+m^Z$ and in parenthesis the
ideal total spin moment predicted by the Slater-Pauling rule for
half-metals (see Ref. \cite{Gala-Full}). With Mn(imp) we denote
the Mn atoms sitting at perfect Co sites.}
 \begin{tabular}{lccccc}
  \hline \hline
& \multicolumn{5}{c}{[Co$_{1-x}$Mn$_x$]$_2$MnAl} \\
 $x$ & Co   & Mn(imp) & Mn & Al & Total(Ideal) \\
  0    &  0.68  & -- & 2.82 & -0.14 & 4.04(4.0) \\

0.025 & 0.71 & -2.63 & 2.82 & -0.14 & 3.92(3.9) \\

0.05 & 0.73 & -2.59 & 2.82 & -0.13 & 3.81(3.8) \\

0.1 & 0.78 & -2.49 & 2.83 & -0.12 & 3.61(3.6) \\

0.2 & 0.84 & -2.23 & 2.85 & -0.09 & 3.20(3.2)\\
\hline

& \multicolumn{5}{c}{[Co$_{1-x}$Mn$_x$]$_2$MnGa} \\
 $x$ & Co   & Mn(imp) & Mn & Ga & Total(Ideal) \\
  0    &  0.65  & -- & 2.90 & -0.10 & 4.09(4.0) \\

0.025 & 0.68 & -2.76 & 2.90 & -0.10 & 4.00(3.9) \\

0.05 & 0.71 & -2.73 & 2.91 & -0.09 & 3.89(3.8) \\

0.1 & 0.75 & -2.66 & 2.92 & -0.08 & 3.66(3.6) \\

0.2 & 0.82 & -2.42 & 2.94 & -0.05 & 3.23(3.2)\\

 \hline \hline
\end{tabular}
\label{table1}
\end{table}

Our discussion has been focused mainly to the half-metallic
character of the compounds with the Mn defects. We will now
concentrate on the spin moments  of the compounds under study to
discuss the appearance of the ferrimagnetism in the defected
compounds. In Table \ref{table1} we have gathered the
atom-resolved and total spin moments for the
[Co$_{1-x}$Mn$_x$]$_2$MnZ alloy where Z is Al or its isoelectronic
Ga and for all values of the concentrations $x$ which we have
studied, and in Table \ref{table2} we present the same information
for the Z= Si, Ge or Sn alloys. We will start our discussion from
the calculated total spin moments and how they compare with the
ideal values for perfect half-metallicity (values in parenthesis
in the tables). These ideal values stem from the Slater-Pauling
behavior shown for the full-Heusler alloys \cite{Gala-Full} which
states that the total spin moment in the unit cell in $\mu_B$ is
the total number of valence electrons in the unit cell, $z_t$,
minus "24" since there are exactly 12 occupied minority-spin
states for all half-metallic full-Heuslers. For the perfect
compounds this means a total spin moment of 4 $\mu_B$ for
Co$_2$MnAl and Co$_2$MnGa since they have 28 valence electrons in
the unit cell and a spin moment of 5 $\mu_B$ for the Co$_2$MnSi,
Co$_2$MnGe and Co$_2$MnSn compounds which have one electron more.
The calculated total spin moments are exactly 5 $\mu_B$ for the
perfect Co$_2$MnSi and Co$_2$MnGe alloys and only slightly deviate
from the ideal values for the other three compounds confirming the
discussion on the total DOS. When we induce the Mn impurities at
the antisites, we have to take the average value of the valence
electrons calculated as $2\times
[(1-x)*z^{Co}+x*z^{Mn(imp)}]+z^{Mn}+z^{sp\:atom}$ where $z$ is the
number of valence electrons of the corresponding chemical element.
Mn is lighter than Co and thus as we increase the concentration in
Mn defects, the total spin moment should decrease. In the case of
the [Co$_{1-x}$Mn$_x$]$_2$Mn-Si and -Ge compounds the ideal
half-metallicity is preserved irrespectively of the concentration
of the Mn impurities. The Al and Sn compounds are almost
half-metallic while slightly larger deviations are observed for
the Ga compound. Overall we could safely state that
half-metallicity is preserved upon the creation of Mn antisites.

\begin{table}
\caption{ Similar to Table \ref{table1} for the
[Co$_{1-x}$Mn$_x$]$_2$MnZ compounds with Z= Si, Ge or Sn.}
 \begin{tabular}{lccccc}
  \hline \hline
& \multicolumn{5}{c}{[Co$_{1-x}$Mn$_x$]$_2$MnSi} \\

$x$ & Co   & Mn(imp) & Mn & Si & Total(Ideal)\\

0.0&  0.98  & -- & 3.13 & -0.09 & 5.00(5.0)\\

0.025 & 0.99& -1.01 & 3.11 & -0.09 & 4.90(4.9) \\

0.05& 0.99 & -0.95 & 3.09 & -0.08 & 4.80(4.8)\\

0.1 & 0.99 & -0.84 & 3.06 & -0.07 & 4.60(4.6) \\

0.2& 0.97 & -0.70 & 2.99 & -0.05 & 4.20(4.2) \\

\hline

& \multicolumn{5}{c}{[Co$_{1-x}$Mn$_x$]$_2$MnGe} \\
$x$ & Co   & Mn(imp) & Mn & Ge & Total(Ideal) \\

0 &  0.93  & -- & 3.20 & -0.06 & 5.00(5.0) \\

0.025 & 0.94& -1.72 & 3.21 & -0.06 & 4.90(4.9) \\

0.05 & 0.95 & -1.63 & 3.20 & -0.05 & 4.80(4.8) \\

0.1 & 0.97 & -1.45 & 3.19 & -0.04 & 4.60(4.6) \\

0.2 & 0.96 & -1.18 & 3.15 & -0.02 & 4.20(4.2) \\ \hline

& \multicolumn{5}{c}{[Co$_{1-x}$Mn$_x$]$_2$MnSn} \\
 $x$ &  Co   & Mn(imp) & Mn & Sn & Total(Ideal) \\
  0   &  0.89  & -- & 3.32 & -0.08 & 5.02(5.0) \\

0.025&  0.91 & -2.44 & 3.34 & -0.07 & 4.91(4.9) \\

0.05&   0.93 & -2.34 & 3.35& -0.06 & 4.81(4.8) \\

0.1&   0.96 & -2.14 & 3.35 & -0.05 & 4.60(4.6) \\

0.2&   0.98 & -1.73 & 3.35 & -0.03 & 4.20(4.2)\\

\hline \hline
\end{tabular}
\label{table2}
\end{table}

The spin moment of Mn atoms at the perfect sites remains
practically constant for all five families of compounds under
study, when the concentration in Mn defects is increased,  since
their environment does not significantly change; each Mn atom has
eight Co atoms as first neighbors in the perfect alloy and for the
values of the concentration under study here their environment
remains of mainly Co character. Co atoms in the case of Si, Ge and
Sn compounds retain a practically constant moment while in Al and
Ga compounds it slightly increases. This implies that in the case
of Si, Ge and Sn-based compounds Co atoms show a more atomic like
behavior and they are not strongly affected by their environment
and thus their spin moment is around 1 $\mu_B$. In the case of Al
and Ga compounds Co atoms are more affected by their environment
and for this reason their spin moment in the perfect compounds is
smaller. As we increase the concentration of Mn defects in these
two alloys, also the spin moment of the Co atoms increases
reaching a value of $\sim$ 0.8 $\mu_B$. Since the spin moments of
both Co and Mn atoms at the perfect sites do not present drastic
changes, the only way to achieve the half-metallicity for the
compounds (and thus to decrease their total spin moment) is that
the Mn impurity atoms at the antisites have spin moments
antiparallel to the other transition metal atoms. This is actually
the case as can be seen in Tables \ref{table1} and \ref{table2}.
Mn impurity atoms have an important negative spin moment
especially in the case of the Al and Ga compounds where it attends
-$\sim$ 2.6-2.8 $\mu_B$. It slightly drops as we increase the
concentration in Mn defects but this is expected since we increase
also the hybridization between Mn impurity atoms; Mn atoms
contrary to Co atoms have more expanded $d$-like wavefunctions and
hybridize stronger between them.

Of particular interest is the case  of Si , Ge and Sn -compounds,
where the spin moment of the Mn impurity atoms shows large
variations between the three families of alloys. As we pass from
Si to Ge and then to Sn-based alloy, its spin moment for $x$=0.025
increases by  -0.72 $\mu_B$. Although the spin moment of the Mn
impurity atom changes significantly, its contribution to the total
spin moment is small. A large increase of the absolute value of
the spin moment of the Mn impurities, as the one observed here,
means a small increase of its negative contribution to the total
spin moment which has to balance the larger positive contribution
of the spin moment of the Mn atoms at the perfect sites observed
for the compound with the heavier sp element. For the Ge- and
Sn-compounds the DOS has a similar shape to the one for the
Si-based alloys (and thus they are not shown here), but the Fermi
level is lower in energy with respect the Si-based compound. Since
the majority DOS takes large values at the region of the minority
gap, this small shift of the Fermi level provokes a considerable
increase of the negative spin moment of the Mn impurity atoms.
Thus the Mn-doped alloys are half-metallic ferrimagnets and their
total spin moment is considerable smaller than the perfect
half-metallic ferromagnetic parent compounds. A similar phenomenon
has been also predicted when creating Cr antisites in the
zinc-blende CrAs intermetallic alloy \cite{Galanakis-RC} where Cr
impurities spin moments couple antiferromagnetically to the Cr
atoms at the ideal sites in the zinc-blende structure. Similarly
Cr antisites in Co$_2$CrAl and Co$_2$CrSi compounds lead to a
similar phenomenon \cite{PSS-RRL}. The importance of the present
results stem from the more intense research on the compounds
containing Co and Mn and from their more robust half-metallic
character, since in the case of Cr compounds the gap almost
vanishes upon the creation of Cr defects.  Here we have to mention
that if also Co atoms migrate to Mn sites (case of atomic swaps)
the half-metallity is lost, as it was shown by Picozzi et al.
\cite{Picozzi04}, due to the energy position of the Co states
which have migrated at Co sites.

\section{Summary and conclusions}
We have studied the effect of defects-driven appearance of
half-metallic ferrimagnetism in the case of the Co$_2$MnZ Heusler
alloys, where Z stands for Al, Ga, Si, Ge or Sn. More precisely,
based on first-principles calculations we have shown that when we
create Mn antisites at the Co sites, these impurity Mn atoms
couple antiferromagnetically with the Co and the Mn atoms at the
perfect sites while keeping the half-metallic character of the
parent compounds. Half-metallicity in these compounds is a robust
property since the large exchange splitting of the Mn impurity
atoms ensures that the width of the gap in the minority spin band
only marginally is affected. Thus we have shown an alternative way
to create robust half-metallic ferrimagnets, which are crucial for
magnetoelectronic applications, based in the introduction of
defects in half-metallic ferromagnetic Heusler alloys which are
widely studied. Especially the case of [Co$_{1-x}$Mn$_x$]$_2$MnSi
alloys is of particular interest since Mn$_3$Si is known to
crystallize in the Heusler $L2_1$ lattice structure of Co$_2$MnSi.
We expect these results to stimulate further interest in the
theoretical and experimental research in the field of spintronics.


\begin{thebibliography}{00}

\bibitem{Zutic}
I. \v{Z}uti\'c, J. Fabian, S. Das Sarma, Rev.
 Mod. Phys. 76 (2004) 323.

\bibitem{book}
Half-metallic alloys: fundamentals and applications, Eds.: I.
Galanakis and P.H. Dederichs, Lecture notes in Physics vol.~676
(Berlin Heidelberg: Springer 2005).


\bibitem{Reviews}
I. Galanakis,  Ph. Mavropoulos, P.H. Dederichs, J. Phys. D: Appl.
Phys. 39 (2006) 765.

\bibitem{deGroot}
R.A. de Groot. F.M. Mueller, P.G. van Engen, K.H.J. Buschow, Phys.
Rev. Lett. 50 (1983) 2024.

\bibitem{Landolt}
P. J. Webster and K. R. A. Ziebeck, in {\em Alloys and Compounds
of d-Elements with Main Group Elements. Part 2.}, edited by H. R.
J. Wijn, Landolt-Bo\"ornstein, New Series, Group III, Vol. 19,Pt.c
(Springer-Verlag, Berlin 1988), pp. 75-184.

\bibitem{Gala-Half}
I. Galanakis, P.H. Dederichs, N. Papanikolaou, Phys. Rev. B 66
(2002) 134428.

\bibitem{Webster}
P.J. Webster,  J. Phys. Chem. Solids 32 (1971) 1221.

\bibitem{Ishida-Fujii}
S. Ishida, S. Fujii, S. Kashiwagi, S. Asano,  J. Phys. Soc. Jpn.
64 (1995) 2152; S. Fujii, S. Sugimura, S. Ishida, S. Asano, J.
Phys.: Condens. Matter 2 (1990) 8583.

\bibitem{Picozzi}
S. Picozzi, A. Continenza, A.J. Freeman, Phys. Rev. B (2002)
094421.

\bibitem{Gala-Full}
I. Galanakis, P.H. Dederichs, N. Papanikolaou, Phys. Rev. B 66
(2002) 174429.

\bibitem{Westerholt}
A. Bergmann, J. Grabis, B.P. Toperverg, V. Leiner, M. Wolff, H.
Zabel, K. Westerholt, Phys. Rev. B 72 (2005) 214403; J. Grabis, A.
Bergmann, A. Nefedov, K. Westerholt, H. Zabel, Phys. Rev. B 72
(2005) 024437; \textit{idem}, Phys. Rev. B 72 (2005) 024438.

\bibitem{Elmers}
A. Rata, H. Braak, D.E. B\"urgler, S. Cramm, and C.M. Schneider,
Eur. Phys. J. B 52 (2006) 445; M. Kallmayer H. Schneider, G.
Jakob, H.J. Elmers, K. Kroth, H.C. Kandpal, U. Stumm, S. Cramm,
Appl. Phys. Lett. 88 (2006) 072506; S.V. Karthik, A. Rajanikanth,
Y.K. Takahashi, T. Okhubo, K. Hono, Appl. Phys. Lett. 89 (2006)
052505; G.H. Fecher, H.C. Kandpal, S. W\"urmehl, J. Morais, H.-J.
Lin, H.-J. Elmers, G. Sch\"onhense,  C. Felser, J. Phys.: Condens.
Matter 17 (2005) 7237; R.Y. Umetsu, K. Kobayashi, A. Fujita, K.
Oikawa, R. Kainuma, K. Ishida, N. Endo, K. Fukamichi, A. Sakuma,
Phys. Rev. B 72 (2005) 214412; K. Kobayashi, R.Y. Umetsu, R.
Kainuma, K. Ishida, T. Oyamada, A. Fujita, K. Fukamichi, Appl.
Phys. Lett. 85 (2004) 4684; H.J. Elmers, G.H. Fecher, D.
Valdaitsev, S.A. Nepijko, A, Gloskovskii, G. Jakob, G.
Sch\"onhense, S. Wurmehl, T. Block, C. Felser, P.-C. Hsu, W.-L.
Tsai, S. Cramm, Phys. Rev. B 67 (2003) 104412.

\bibitem{Marukame}
T. Marukame,  T. Ishikawa, K.-I. Matsuda, T. Uemura, M. Yamamoto,
Appl. Phys. Lett. 88 (2006) 262503; T. Marukame, T. Kasahara, K.
Matsuda, T. Uemura T M. Yamamoto, Jpn. J. Appl. Phys. 44 (2005)
L521.

\bibitem{Kelekar}
R. Kelekar and B.M. Klemens, Appl. Phys. Lett. 86 (2005) 232501;
R. Kelekar, H. Ohldag,  B.M. Clemens, Phys. Rev. B 75 (2007)
014429.

\bibitem{Reiss}
S. K\"ammerer, A. Thomas, A. H\"utten, G. Reiss,  Appl. Phys.
Lett. 85 (2004) 79;  J. Schmalhorst, S. K\"ammerer, M. Sacher, G.
Reiss,  A. H\"utten,  A. Scholl, Phys. Rev. B 70 (2004) 024426.

\bibitem{Sakuraba}
Y. Sakuraba, M. Hattori, M. Oogane, Y. Ando, H. Kato, A. Sakuma,
T. Miyazaki, H. Kubota,  Appl. Phys. Lett. 88 (2006) 192508; Y.
Sakuraba, J. Nakata, M. Oogane, Y. Ando, H. Kato, A. Sakuma, T.
Miyazaki, H. Kubota,  Appl. Phys. Lett. 88 (2006)022503; Y.
Sakuraba, T. Miyakoshi, M. Oogane, Y. Ando, A. Sakuma, T.
Miyazaki, H. Kubota,  Appl. Phys. Lett. 89 (2006) 052508; X.Y.
Dong, C. Adelmann, J.Q. Xie, C.J. Palmstr\"om, X. Lou, J. Strand,
P.A. Crowell, J.-P. Barnes, A.K. Petford-Long, Appl. Phys. Lett.
86 (2005) 102107.

\bibitem{magnetism}
W.H. Wang, M. Przybylski, W. Kuch, L.I. Chelaru, J. Wang, Y.F. Lu,
J. Barthel, H.L. Meyerheim, J. Kirschner, Phys. Rev. B 71 (2005)
144416; S. Wurmehl, G.H. Fecher, H.C. Kandpal, V. Ksenofontov, C.
Felser, H.-J. Lin, J. Morais, Phys. Rev. B 72 (2005) 184434; R.Y.
Umetsu, K.Kobayashi, R.Kainuma, A. Fujita, K. Fukamichi, K.Ishida,
A. Sakuma, Appl. Phys. Lett. 85 (2004) 2011; L. Ritchie, G. Xiao,
Y. Ji, T.Y. Chen, C.L. Chien, M. Zhang, J. Chen, Z. Liu, G. Wu,
X.X. Zhang, Phys. Rev.  B 68 (2003) 104430.

\bibitem{superlattices}
K. Yakushiji,  K. Saito, S. Mitani, K. Takanashi, Y.K. Takahashi,
K. Hono, Appl. Phys. Lett. 88 (2006) 222504; M. Hashimoto, J.
Herfort, H.-P. Sch\"onherr, K.H. Ploog, Appl. Phys. Lett. 87
(2005) 102506.

\bibitem{transport}
N.-N. Liu, A. Thomas, G. Reiss, A. H\"utten, Appl. Phys. Lett. 89
(2006) 162506; Z. Gercsi, A. Rajanikanth, Y.K. Takahashi, K. Hono,
M. Kikuchi, N. Tezuka, K. Inomata,  Appl. Phys. Lett. 89 (2006)
082512.

\bibitem{devices}
N. Tezuka, N. Ikeda, A. Miyazaki, S. Sugimoto, M. Kikuchi, K.
Inomata, Appl. Phys. Lett.  89 (2006) 112514; S. Okamura, A.
Miyazaki, S. Sugimoto, N. Tezuka, K. Inomata, Appl. Phys. Lett.
86(2005) 232503.


\bibitem{Sasioglu}
E. \c Sa\c s\i o\~glu, L. M. Sandratskii, P. Bruno, and I.
Galanakis, Phys. Rev. B \textbf{72}, 184415 (2005).

\bibitem{Leuken}
H. van Leuken and R.A. de Groot, Phys. Rev. Lett. 74 (1995) 1171;
S. Wurmehl, H.C. Kandpal, G.H. Fecher, C. Felser, J. Phys.:
Condens. Matter 18 (2006) 6171.

\bibitem{Groot2}
R.A. de Groot, A.M. van der Kraan, K.H.J. Buschow, J. Magn. Magn.
Mater. 61 (1986) 330.

\bibitem{Kemal}
K. \"Ozdo\~gan, I. Galanakis, E. \c Sa\c s\i o\~glu, B. Akta\c s,
J. Phys.: Condens. Matter  18 (2006) 2905; E. \c Sa\c s\i o\~glu,
L.M. Sandratskii,  P. Bruno, J. Phys.: Condens. Matter 17 (2005)
995.

\bibitem{Akai}
H. Akai and M. Ogura, Phys. Rev. Lett. 97 (2006) 026401.

\bibitem{Galanakis-RC}
I. Galanakis, K. \"Ozdo\~gan,  E. \c Sa\c s\i o\~glu, B. Akta\c s,
Phys. Rev. B 74 (2006) 140408(R).

\bibitem{PSS-RRL}
K.  \"Ozdo\~gan, I. Galanakis, E. \c Sa\c s\i o\~glu, B. Akta\c s,
Phys. Stat. Sol. (RRL) 1 (2007) R95.

\bibitem{Podlucky}
X.-Q. Chen, R. Podloucky, P. Rogl, J. Appl. Phys. 100 (2006)
113901.

\bibitem{Richter}
M. Sargolzaei, M. Richter, K. Koepernik, I. Opahle, H. Eschrig, I.
Chaplygin, Phys. Rev. B 74 (2006) 224410.

\bibitem{JAP}
K.  \"Ozdo\~gan, B. Akta\c s, I. Galanakis, E. \c Sa\c s\i o\~glu,
J. Appl. Phys. in press [preprint cond-mat/0612194]

\bibitem{APL}
I. Galanakis, K.  \"Ozdo\~gan,  B. Akta\c s,  E. \c Sa\c s\i
o\~glu, Appl. Phys. Lett. 89 (2006) 042502; K.  \"Ozdo\~gan,  E.
\c Sa\c s\i o\~glu, B. Akta\c s, I. Galanakis, Phys. Rev. B 74
(2006) 172412; I. Galanakis, J. Phys.: Condens. Matter 16 (2004)
3089.

\bibitem{different}
V.N. Antonov, H.A. D\"urr, Yu. Kucherenko, L.V. Bekenov, A.N.
Yaresko, Phys. Rev B 72 (2005) 054441; Y. Miura, K. Nagao,  M.
Shirai,  Phys. Rev B 69 (2004) 144413; B. Balke, G.H. Fecher, H.C.
Kandpal, C. Felser, K. Kobayashi, E. Ikenaga, J.-J. Kim, S. Ueda,
Phys. Rev. B 74 (2006) 104405.

\bibitem{Mn3Si}
C. Pfleiderer, J. Boeuf, H.v. L\"ohneysen, Phys. Rev. B 65 (2002)
172404.

\bibitem{Mn3Ge}
J.W. Cable, N. Wakabayashi, P. Radhakrishna, Phys. Rev. B 48
(1993) 6159.


\bibitem{Picozzi04}
S. Picozzi, A. Continenza, A.J. Freeman,   Phys. Rev. B 69 (2004)
094423.

\bibitem{koepernik}
K. Koepernik and H. Eschrig, Phys. Rev. B 59 (1999) 3174; K.
Koepernik, B. Velicky, R. Hayn, H. Eschrig, Phys. Rev. B 58 (1998)
6944.

\end{thebibliography}
\end{document}